\documentclass[pre,amsmath, amsfonts,twocolumn, superscriptaddress]{revtex4}
\usepackage{graphicx}
\usepackage{braket}
\usepackage[normalem]{ulem}
\usepackage{xspace}

\usepackage[dvipsnames]{xcolor}
\usepackage{subfigure}
\begin{document}

\title{Reply: Inferring entropy production from waiting time distributions}

\author{Gili Bisker} 
\email{bisker@tauex.tau.ac.il} 
\affiliation{Department of Biomedical Engineering, Faculty of Engineering, Tel Aviv University, Tel Aviv 6997801, Israel}
\affiliation{Center for Physics and Chemistry of Living Systems, Center for Nanoscience and Nanotechnology, Center for Light-Matter Interaction, Tel Aviv University, Tel Aviv 6997801, Israel}
\author{Ignacio A. Mart\'inez}
\email{ignacio.martinez@ugent.be} 
\affiliation{Electronics and Information Systems, Ghent University, Technologiepark Zwijnaarde 15, 9052 Gent, Belgium}
\author{Jordan M. Horowitz}
\email{jmhorow@umich.edu} 
\affiliation{Department of Biophysics, University of Michigan, Ann Arbor, Michigan, 48109, USA}
\affiliation{Center for the Study of Complex Systems, University of Michigan, Ann Arbor, Michigan 48104, USA}
\affiliation{Department of Physics, University of Michigan, Ann Arbor, Michigan, 48109, USA}
\author{Juan M.R. Parrondo}
\email{parrondo@ucm.es} 
\affiliation{Departamento de Estructura de la Materia, F\'isica Termica y Electronica and GISC, Universidad Complutense de Madrid 28040 Madrid, Spain}

\date{\today}

\date\today
%\begin{abstract}
%We discuss the criticism 
%\end{abstract}

\maketitle

In \cite{ugly},
Hartich and Godec (HG) present a counterexample that apparently refutes our results in \cite{nice}. 
{While the} content of their Comment is essentially correct, it does not invalidate our results, but rather 
raises an interesting question on the effect of coarse-graining on irreversibility, which we discuss in detail below.
However, the way they write the Comment is slightly misleading. They highlight the fact that we never tested Eq.~(4) in Ref.~\cite{nice}, implicitly suggesting that this equation is wrong.
Eqs.~(2-4) are an exact expression of the Kullbak-Leibler divergence (KLD) between a semi-Markov chain and its time reverse. There is no need of a ``test'' since it is an exact result, whose  rigorous mathematical proof is given in the section {\em Methods} of Ref.~\cite{nice}.
Moreover, HG  claim that our results and those of Wang and Qian \cite{wang} are ``diametrically opposing''. We want to stress that this is not true. Both our paper and  \cite{wang} are fully correct and perfectly compatible.

\begin{figure}
\centering
\includegraphics[scale=0.4]{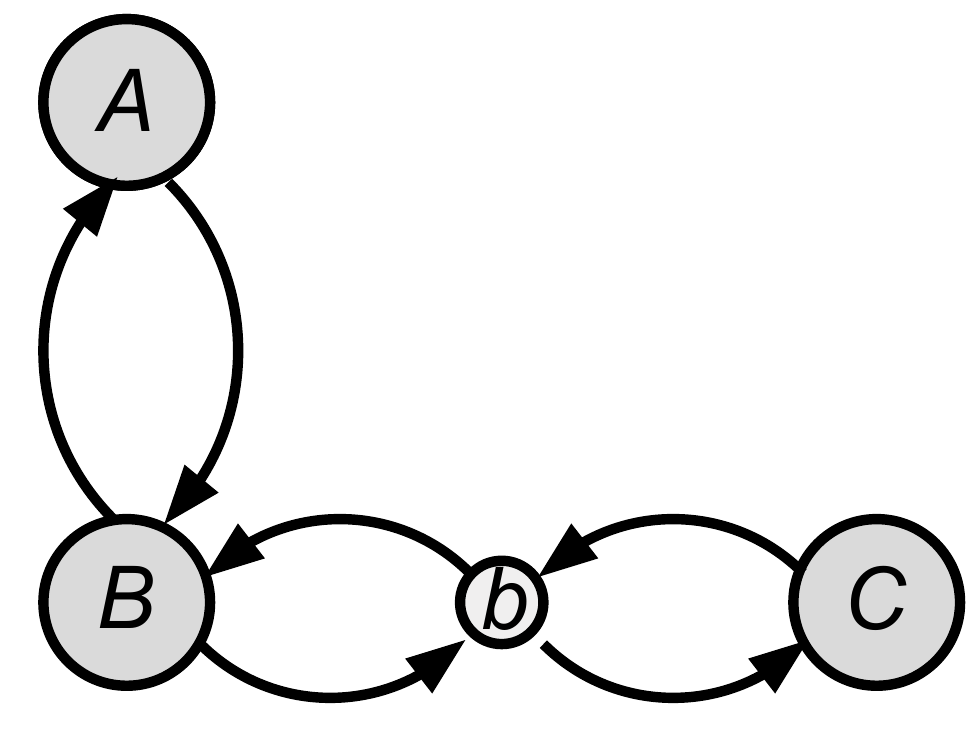}\\
\includegraphics[width=0.80\columnwidth]{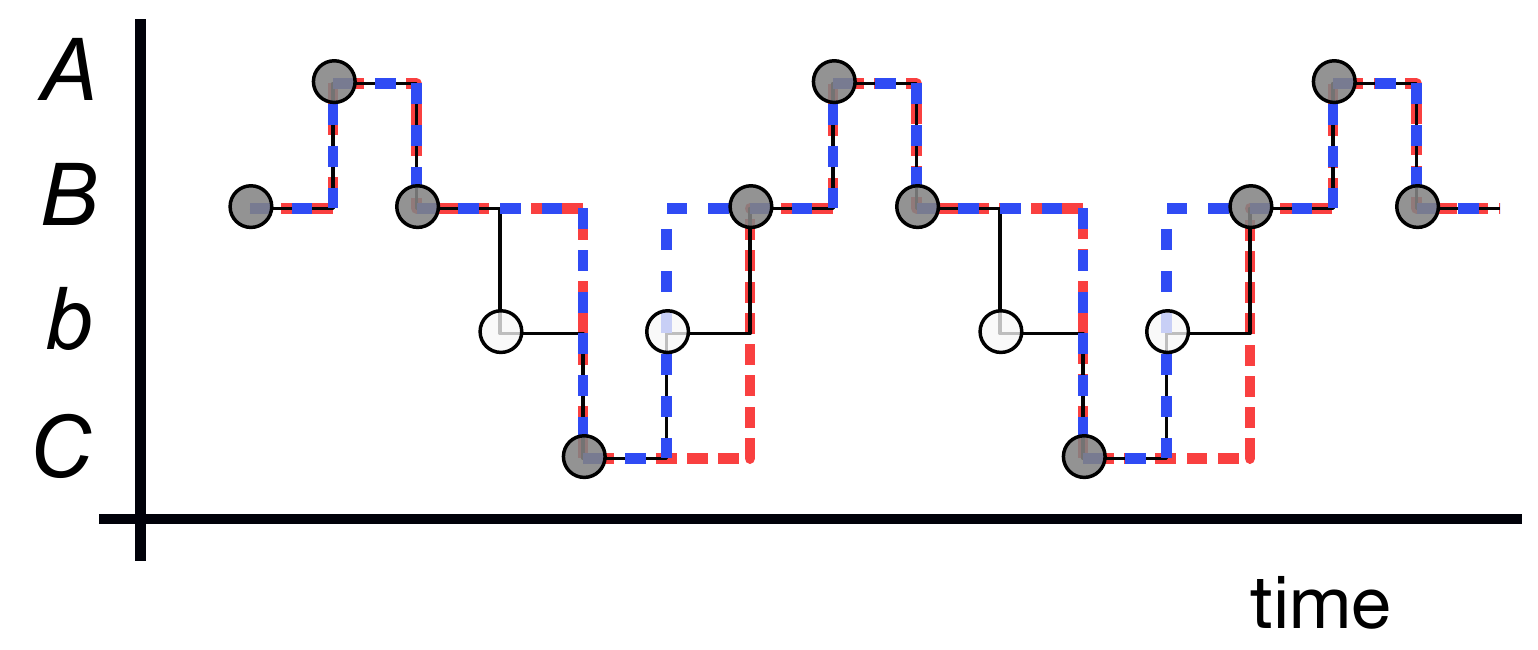}
\caption{
{\em Top)} A simplified version of the counterexample introduced in \cite{ugly}. We decimate state $b$.  {\em Bottom)} We plot microscopic trajectory (thin black lines) and the  trajectories (dashed lines) resulting from different decimation procedures, where state $b$ (open circles) is not observable. The blue dashed line is the result of lumping states $B$ and $b$. Since the decimation is local in time, decimation and time reversal commute, $f(\Theta X)=\Theta f(X)$. Here, the microscopic trajectory $X$ is reversible, hence $\Theta  X=X$, and the commutation implies that $f(X)$ is also reversible, as shown in the figure.   On the other hand, the red dashed line results from the decimation procedure considered in \cite{ugly}, which is based on the arrivals at states $A$ and $B$. This decimation does not commute with the time-reversal operation, and the resulting trajectory is irreversible $\Theta f(X)\neq f(X)=f(\Theta X)$: in the red trajectory the system takes two units of time to jump down from $B$ to $C$ and one unit of time to jump up from $B$ to $A$; whereas these waiting times are swapped when the trajectory is reversed.  Notice that this decimation procedure is non local in time, i.e., the state of the decimated trajectory at time $t$ is not a function of the micro-state at time $t$ (open circles),  but depends on the past.}
\label{fig:threestates}
\end{figure}

The example in \cite{ugly} has nothing to do with the validity of Eq.~(4). Instead, it calls into
question Eq.~(1) in our paper \cite{nice}, which asserts that the KLD between an observed trajectory and its time reversal is a lower bound to the {physical } entropy production. This claim is widely used in the literature and based on a well-known property of the KLD. The KLD between two stochastic processes measures our capacity to distinguish {between} them using data. If we remove information about these processes, 
{by} adding noise to the data or decimating the underlying network, it should be clear that this capacity decreases and, consequently, the KLD decreases. Since the entropy production is equal to the KLD  between the microscopic trajectory and its time reversal \cite{kawai,edgar,edgar2}, the KLD between a coarse-grained trajectory and its time reversal is a lower bound
{on} the actual entropy production.

In more precise mathematical terms, this argument is based on the following property of the KLD between two distributions $p_{X}$ and $q_{X}$ of a random variable $X$:
\begin{equation}\label{ine}
D(p_{X}||q_{X})\geq D(p_{f(X)}||q_{f(X)})
\end{equation}
where $f(x)$ is an arbitrary {(possibly random)} function. The equality holds if $f$ is {one-to-one and deterministic}. The function $f$ can represent a coarse-graining when several states $\{x_{1},\dots,x_{m}\}$ coalesce
into a single state $f(x_{1})=\dots =f(x_{m})$, or the removal of the information associated to a {particular} variable, when for instance $x=(x^{(1)},x^{(2)})$ and $f(x^{(1)},x^{(2)})=x^{(1)}$.

The inequality \eqref{ine} has a direct and easy interpretation: by manipulating the available data, that is, by transforming the data according to the function $f$, one cannot increase the distinguishability between the two probability distributions, $p$ and $q$.

If $X$ is a microscopic trajectory and $\Theta X$ is its time reversal, the entropy production $\Delta S$ verifies \cite{kawai,edgar} 
\begin{equation}\label{ine2}
\Delta S=kD(p_{X}||p_{\Theta X}).
\end{equation}
where $k$ is Boltzmann constant. If we apply Eq.~\eqref{ine} to this expression, we obtain 
\begin{equation}\label{ine3}
\Delta S\geq kD(p_{f(X)}||p_{f(\Theta X)}).
\end{equation}
However, Eq.~(1) in our work \cite{nice} slightly differs from this inequality. There, we assume that the observer has access to a coarse-grained trajectory $f(X)$ and constructs the time reversal as $\Theta f(X)$. Hence, to have
\begin{equation}\label{ine4}
\Delta S\geq kD(p_{f(X)}||p_{\Theta f(X)})
\end{equation}
from Eq.~\eqref{ine3},
a sufficient condition is that coarse-graining and time reversal commute, $f(\Theta X)=\Theta f(X)$.

The example in \cite{ugly} does not fulfill this commutation relation, as HG mention in their comment.
In Fig.~\ref{fig:threestates},
we introduce a simplified version of the counterexample discussed in \cite{ugly} that illustrates the origin of this issue. The example is a kinetic network at equilibrium with four states, $A$, $B$, $C$, and $b$. Suppose that the observer does not have access to state $b$. There are different options for constructing the coarse-grained trajectory. The one discussed by HG consists of taking the arrivals at the states $A$, $B$, and $C$ as  jumps between states in the coarse-grained trajectory. 
In Fig.~\ref{fig:threestates} ({\em bottom}), we plot a sketch of reversible microscopic trajectory (thin black lines) $X$,  whereas the corresponding coarse-grained trajectory using this prescription (red dashed lines). Since $X$ is reversible, $X=\Theta X$, and the resulting trajectory is irreversible, hence $\Theta f(X)\neq f(X)=f(\Theta X)$. The irreversibility of  the red trajectory is revealed by the waiting time distributions: in the forward trajectory, $f(X)$, the system takes two units of time to jump down from $B$ to $C$ and one unit of time to jump up from $B$ to $A$, whereas these waiting times are swapped when the trajectory is reversed.  Notice that this decimation procedure is non local in time, i.e., the state of the decimated trajectory at time $t$ is not a function of the micro-state at time $t$ (open circles),  but depends on the past.

We also plot in Fig.~\ref{fig:threestates} ({\em bottom}), the trajectory  resulting from lumping states $B$ and $b$ (blue dashed lines), which is reversible. This procedure is local in time, since the hidden state $b$ is always mapped onto $B$, and consequently decimation and time reversal commute, $f(\Theta X)=\Theta f(X)$.

Summarizing, HG comment is interesting since it points out that a sufficient condition for Eq.~(1) to be valid is that coarse-graining and time reversal commute. 
This condition was not mentioned in our paper because we implicitly assumed coarse-graining procedures that are local in time and consequently commute with the time-reversal operation, as implied in the description of the decimation process in the main text and in the {\em Methods} section in \cite{nice}.

\bibliography{rep.bib}

\begin{thebibliography}{6}%
\makeatletter
\providecommand \@ifxundefined [1]{%
 \@ifx{#1\undefined}
}%
\providecommand \@ifnum [1]{%
 \ifnum #1\expandafter \@firstoftwo
 \else \expandafter \@secondoftwo
 \fi
}%
\providecommand \@ifx [1]{%
 \ifx #1\expandafter \@firstoftwo
 \else \expandafter \@secondoftwo
 \fi
}%
\providecommand \natexlab [1]{#1}%
\providecommand \enquote  [1]{``#1''}%
\providecommand \bibnamefont  [1]{#1}%
\providecommand \bibfnamefont [1]{#1}%
\providecommand \citenamefont [1]{#1}%
\providecommand \href@noop [0]{\@secondoftwo}%
\providecommand \href [0]{\begingroup \@sanitize@url \@href}%
\providecommand \@href[1]{\@@startlink{#1}\@@href}%
\providecommand \@@href[1]{\endgroup#1\@@endlink}%
\providecommand \@sanitize@url [0]{\catcode `\\12\catcode `\$12\catcode
  `\&12\catcode `\#12\catcode `\^12\catcode `\_12\catcode `\%12\relax}%
\providecommand \@@startlink[1]{}%
\providecommand \@@endlink[0]{}%
\providecommand \url  [0]{\begingroup\@sanitize@url \@url }%
\providecommand \@url [1]{\endgroup\@href {#1}{\urlprefix }}%
\providecommand \urlprefix  [0]{URL }%
\providecommand \Eprint [0]{\href }%
\providecommand \doibase [0]{http://dx.doi.org/}%
\providecommand \selectlanguage [0]{\@gobble}%
\providecommand \bibinfo  [0]{\@secondoftwo}%
\providecommand \bibfield  [0]{\@secondoftwo}%
\providecommand \translation [1]{[#1]}%
\providecommand \BibitemOpen [0]{}%
\providecommand \bibitemStop [0]{}%
\providecommand \bibitemNoStop [0]{.\EOS\space}%
\providecommand \EOS [0]{\spacefactor3000\relax}%
\providecommand \BibitemShut  [1]{\csname bibitem#1\endcsname}%
\let\auto@bib@innerbib\@empty
%</preamble>
\bibitem [{\citenamefont {Hartich}\ and\ \citenamefont {Godec}(2021)}]{ugly}%
  \BibitemOpen
  \bibfield  {author} {\bibinfo {author} {\bibfnamefont {D.}~\bibnamefont
  {Hartich}}\ and\ \bibinfo {author} {\bibfnamefont {A.}~\bibnamefont
  {Godec}},\ }\href@noop {} {\  (\bibinfo {year} {2021})},\ \Eprint
  {http://arxiv.org/abs/2112.08978} {arXiv:2112.08978 [cond-mat.stat-mech]}
  \BibitemShut {NoStop}%
\bibitem [{\citenamefont {Mart\'inez}\ \emph {et~al.}(2019)\citenamefont
  {Mart\'inez}, \citenamefont {Bisker}, \citenamefont {Horowitz},\ and\
  \citenamefont {Parrondo}}]{nice}%
  \BibitemOpen
  \bibfield  {author} {\bibinfo {author} {\bibfnamefont {I.~A.}\ \bibnamefont
  {Mart\'inez}}, \bibinfo {author} {\bibfnamefont {G.}~\bibnamefont {Bisker}},
  \bibinfo {author} {\bibfnamefont {J.~M.}\ \bibnamefont {Horowitz}}, \ and\
  \bibinfo {author} {\bibfnamefont {J.~M.~R.}\ \bibnamefont {Parrondo}},\
  }\href@noop {} {\bibfield  {journal} {\bibinfo  {journal} {Nature
  Communications}\ }\textbf {\bibinfo {volume} {10}},\ \bibinfo {pages} {3542}
  (\bibinfo {year} {2019})}\BibitemShut {NoStop}%
\bibitem [{\citenamefont {Wang}\ and\ \citenamefont {Qian}(2007)}]{wang}%
  \BibitemOpen
  \bibfield  {author} {\bibinfo {author} {\bibfnamefont {H.}~\bibnamefont
  {Wang}}\ and\ \bibinfo {author} {\bibfnamefont {H.}~\bibnamefont {Qian}},\
  }\href@noop {} {\bibfield  {journal} {\bibinfo  {journal} {Journal of
  Mathematical Physics}\ }\textbf {\bibinfo {volume} {48}},\ \bibinfo {pages}
  {013303} (\bibinfo {year} {2007})}\BibitemShut {NoStop}%
\bibitem [{\citenamefont {Kawai}\ \emph {et~al.}(2007)\citenamefont {Kawai},
  \citenamefont {Parrondo},\ and\ \citenamefont {den Broeck}}]{kawai}%
  \BibitemOpen
  \bibfield  {author} {\bibinfo {author} {\bibfnamefont {R.}~\bibnamefont
  {Kawai}}, \bibinfo {author} {\bibfnamefont {J.~M.~R.}\ \bibnamefont
  {Parrondo}}, \ and\ \bibinfo {author} {\bibfnamefont {C.~V.}\ \bibnamefont
  {den Broeck}},\ }\href@noop {} {\bibfield  {journal} {\bibinfo  {journal}
  {Phys. Rev. Lett.}\ }\textbf {\bibinfo {volume} {98}},\ \bibinfo {pages}
  {080602} (\bibinfo {year} {2007})}\BibitemShut {NoStop}%
\bibitem [{\citenamefont {Rold\'an}\ and\ \citenamefont
  {Parrondo}(2010)}]{edgar}%
  \BibitemOpen
  \bibfield  {author} {\bibinfo {author} {\bibfnamefont {E.}~\bibnamefont
  {Rold\'an}}\ and\ \bibinfo {author} {\bibfnamefont {J.~M.~R.}\ \bibnamefont
  {Parrondo}},\ }\href@noop {} {\bibfield  {journal} {\bibinfo  {journal}
  {Phys. Rev. Lett.}\ }\textbf {\bibinfo {volume} {105}},\ \bibinfo {pages}
  {150607} (\bibinfo {year} {2010})}\BibitemShut {NoStop}%
\bibitem [{\citenamefont {Rold{\'{a}}n}\ \emph {et~al.}(2021)\citenamefont
  {Rold{\'{a}}n}, \citenamefont {Barral}, \citenamefont {Martin}, \citenamefont
  {Parrondo},\ and\ \citenamefont {J\"ulicher}}]{edgar2}%
  \BibitemOpen
  \bibfield  {author} {\bibinfo {author} {\bibfnamefont {{\'{E}}.}~\bibnamefont
  {Rold{\'{a}}n}}, \bibinfo {author} {\bibfnamefont {J.}~\bibnamefont
  {Barral}}, \bibinfo {author} {\bibfnamefont {P.}~\bibnamefont {Martin}},
  \bibinfo {author} {\bibfnamefont {J.~M.~R.}\ \bibnamefont {Parrondo}}, \ and\
  \bibinfo {author} {\bibfnamefont {F.}~\bibnamefont {J\"ulicher}},\
  }\href@noop {} {\bibfield  {journal} {\bibinfo  {journal} {New Journal of
  Physics}\ }\textbf {\bibinfo {volume} {23}},\ \bibinfo {pages} {083013}
  (\bibinfo {year} {2021})}\BibitemShut {NoStop}%
\end{thebibliography}%
\bibliographystyle{apsrev4-1}

\end{document}